\documentstyle[11pt,aaspp4,flushrt]{article}  
\def\be{\begin{equation}}
\def\ee{\end{equation}}

\def\msun{M_{\odot}}

\def\Ti{T_{{\rm out},i}}
\def\Te{T_{{\rm out},e}}
\def\gs{\rm \,g\,s^{-1}}

\input psfig.tex
\begin{document}
\date{}
\title{The role of the outer boundary condition in accretion disk models:
theory and application}

\author{Feng Yuan}
\affil{Department of Astronomy, Nanjing University, Nanjing 210093,\\
and Beijing Astrophysics Center, Beijing 100080, China
\\  Email: fyuan@nju.edu.cn}
 
\author{Qiuhe Peng}
\affil{Department of Astronomy, Nanjing University, Nanjing 210093, China}

\author{Ju-fu Lu}
\affil{Center for Astrophysics, University of Science \& Technology of China, 
Hefei, 230026, China,\\ and
National Astronomical Observatories, Chinese Academy of Sciences}

\author{Jianmin Wang}
\affil{Laboratory of Cosmic Ray and High Energy Astrophysics,\\
Institute of High Energy Physics, Chinese Academy of Sciences, Beijing 100039,
 China}

\begin{abstract}
In a previous paper (Yuan 1999, hereafter Paper I), 
we find that the outer boundary conditions(OBCs) of an 
optically thin accretion flow
play an important role in determining
the structure of the flow. Here in this paper, we further
investigate  the influence of OBC
on the dynamics and radiation of the accretion flow on a 
more detailed level. Bremsstrahlung and synchrotron
radiations amplified by Comptonization are 
taken into account and two-temperature
plasma assumption is adopted. The three OBCs we adopted are the 
temperatures of the electrons and ions and the specific angular
momentum of the accretion flow at a certain outer boundary.
We investigate the individual 
role of each of the three OBCs
on the dynamical structure and the emergent spectrum. 
We find that when the general 
parameters such as the mass accretion rate $\dot{M}$ 
and the viscous parameter $\alpha$ are fixed,
the peak flux at various bands such as radio, IR and X-ray,
can differ by as large as 
several orders of magnitude under different OBCs in our 
example. Our results indicate
that OBC is both dynamically and radiatively important
therefore should be regarded as a new ``parameter'' in accretion disk models.

As an illustrative example, we further apply the above 
results to the compact radio source Sgr A$^*$ 
located at the center of our Galaxy. 
Advection-dominated accretion flow (ADAF) model has been turned out to
be a great success to explain its luminosity and spectrum. However,
there exists a discrepancy between the mass accretion rate 
favored by ADAF models in the literature 
and that favored by the three dimensional hydrodynamical simulation,
with the former being 
10-20 times smaller than the latter. 
By seriously considering the
outer boundary condition of the accretion flow, we find that
due to the low specific angular momentum of the accretion gas, the
accretion in Sgr A$^*$ should belong to a new accretion pattern,
which is characterized by possessing a very large sonic radius.
This accretion pattern can significantly reduce the discrepancy between 
the mass accretion rate. We argue that
the accretion occurred in some detached binary systems,
the core of nearby elliptical galaxies and active galactic nuclei(AGNs)
very possibly belongs to this accretion pattern.

\end{abstract}

 
\section{INTRODUCTION}
The dynamics of an accretion flow around 
a black hole is described by a set
of non-linear differential equations. 
So this is intrinsically an initial value problem, and generally
the outer boundary condition(OBC) 
has a significant influence on the global solution.
In the standard thin disk model (Shakura \& Sunyaev 1973), all the differential
terms are neglected therefore the equations are reduced to 
a set of algebraic equations which don't entail any boundary condition 
at all. However, in many cases it has been shown that 
the differential terms, such as the inertial and the horizontal pressure
gradient terms in
the radial momentum equation and the advection term in the 
energy equation, play an important role and can not 
be neglected (Begelman 1978; Begelman \&
Meier 1982; Abramowicz et al. 1988). This is especially the case  for an
optically thin advection dominated 
accretion flow (ADAF) (Ichimaru 1977; Rees et al. 1982; Narayan \&
 Yi 1994; Abramowicz et al. 1995; Narayan, Kato \& Honma
 1997; Chen, Abramowicz \& Lasota 1997). It is the existence
of the differential terms that will make OBC an important
factor to determine the behavior of an accretion flow.

This expectation has been initially confirmed in our Paper I.
In that paper, taking optically thin one- and two-temperature
plasma as examples, we investigated the influence of OBC 
on the dynamics of an accretion flow. We adopted 
the temperature and
the ratio of the radial velocity to the local sound speed
(or, equivalently, the angular velocity $\Omega_{\rm out}$) at a certain outer 
boundary $r_{\rm out}$ as the outer boundary conditions 
and found that in 
both cases, the topological structure and the profiles of angular momentum 
and surface density of the flow differ greatly under 
different OBCs. In terms of the topological structure and the profile 
of the angular momentum, three types of solutions are found. 
When $T_{\rm out}$ is relatively low,
the solution is of type I. When $T_{\rm out}$ is relatively high and 
the angular velocity $\Omega_{\rm out}$ is high,
the solution is of type II. Both types I and II possess small sonic radii,
but their topological structures and angular momentum profiles 
are different. When $T_{\rm out}$ is relatively high and the 
angular velocity is lower than a critical value, the solution is of type III,
characterized by possessing a much larger sonic radius. 
For a one-temperature plasma
or ions in a two-temperature plasma, the discrepancy of the temperature
in the solutions with different OBCs lessen rapidly away from the outer
boundary, but the discrepancy of the electron temperature at $r_{\rm out}$
persists throughout the disk. This is because when the  
accretion rate is low, the electrons are basically adiabatic,
i.e., both the local radiative cooling and the energy transfer from
the ions to the electrons can be neglected compared with 
the energy advection of the electrons. While for a one-temperature plasma
or ions, the local viscous dissipation in the energy
balance plays an important role, thus their temperature 
is more {\em locally} determined. 

In the present paper we will focus on the two-temperature
accretion flows. Our first aim is to extend our study in Paper I by including 
synchrotron radiation as well. This is a 
very strong local radiative cooling mechanism 
in the inner region of a disk when a magnetic field is present. We 
will check whether our conclusion of Paper I still holds or not 
in this case.

The main aim of this paper is to calculate the emergent spectrum
under different OBCs and to investigate the applications. We want to
probe whether and how the spectrum is 
dependent upon the OBCs (Sections 2 \& 3). We find that each of the three OBCs 
has a significant influence on the emergent spectrum.
This result indicates the importance of
considering the initial physical state of the accretion flow.
In Section 4 we apply our theory of OBC to Sgr A$^*$. We find that the 
discrepancy between the mass accretion rate favored by
the numerical simulation and that required in all ADAF models in the literature
is naturally solved by seriously considering the OBCs of accretion flows.
The last section is devoted to summary and discussion, where 
the promising applications in nearby elliptical galaxies and 
Active Galactic Nuclei (AGNs) are shortly  discussed.

\section{MODEL}

We consider a steady axisymmetric accretion flow around a Schwarzschild black
hole of mass $M$.
Paczy\'nski \& Wiita (1980) potential $\phi=-GM/(r-r_g)$ is used to mimic the
geometry of the hole, where $r_g=2GM/c^2$ is the Schwarzschild radius. 
The standard $\alpha$-viscosity prescription is adopted. We assume that
all the viscous dissipated energy is transfered to ions and the energy
transfer from ions to electrons is provided solely by Coulomb collisions.
So the plasma has a two-temperature structure 
in the present optically thin case. 
A randomly oriented magnetic field is assumed to exist in 
the accretion flow. The total pressure is then taken to be
\be
p=p_{\rm gas}+p_{\rm mag},
\ee
where $p_{\rm gas}$ is the gas pressure and $p_{\rm mag}$ is the magnetic
pressure. For simplicity, we assume the ratio of gas pressure
to the total pressure $\beta$ to be a global parameter independent of radius
$r$. Under the optically thin assumption, the equation of state
can be written as:
\be
p_{\rm gas}=\beta p=p_i+p_e=\frac{\rho}{\mu_i}\frac{k}{m_
\mu}T_i+\frac{\rho}{\mu_e}\frac{k}{m_\mu}T_e,
\ee
and
\be
p_{\rm mag}=(1-\beta)p=\frac{B^2}{8\pi}.
\ee
Here and hereafter subscripts $i$ and $e$ indicate the quantities for ions and
electrons, respectively. The mean molecular weight for ions and
electrons are:
\be
\mu_i=1.23 \hspace{1cm} \mu_e=1.13,
\ee
respectively. We take the shear stress to be simply 
proportional to the pressure, i.e.,
shear stress=-$\alpha p$. The hydrostatic balance in the vertical direction
is also assumed.

Under the above assumptions, the set of height-integrated  equations describing 
the behavior of accretion flows read as follows.
\be
-4\pi r H\rho v=\dot{M},\hspace{4mm}{\rm with} \hspace{3mm} 
 H=c_s/\Omega_{\rm k} 
\equiv \sqrt{p/\rho}/\Omega_{\rm k},
\ee
\be
v \frac{dv}{dr}=-\Omega_{\rm k}^2 r+\Omega^2 r-\frac{1}{\rho}\frac
{dp}{dr},
\ee
\be
v(\Omega r^2-j)=\alpha r \frac{p}{\rho},
\ee
\be
\rho v \left(\frac{d \varepsilon_i}{dr}+p_i \frac{d}{dr}
 \left( \frac{1}{\rho}\right)
\right)=q^+-q_{ie}=-\alpha p r \frac{d\Omega}{dr}-q_{ie},
\ee
\be
\rho v \left(\frac{d \varepsilon_e}{dr}+p_e \frac{d}{dr}
 \left( \frac{1}{\rho}\right) \right)=q_{ie}-q^-.
\ee
All above quantities have their popular meanings. 
The difference between eq. (7) above and eq. (2.11) in Narayan, 
Kato, \& Honma (1997) or eq. (2.13) in Narayan, Mahadevan, \& Quataert (1998)
is due to the difference of the adopted viscosity prescription
(see, e.g., Abramowicz et al. 1988; Nakamura et al. 1997; Manmoto,
Mineshige, \& Kusunose 1997). We adopt this kind of viscosity prescription
because in this case the no-torque condition at the hole horizon 
is automatically satisfied therefore the calculation can be simplified
(Abramowicz et al. 1988; Narayan, Kato, \& Honma 1997).
$q_{ie}$ denotes the energy transfer rate from ions to electrons by 
Coulomb collisions which takes the form
(Dermer, Liang \& Canfield 1991):
\be
q_{ie}=\frac{3}{2}\frac{m_e}{m_i}n_en_i \sigma_T c {\rm ln} \Lambda
\left(kT_i-kT_e\right)\frac{\left(\frac{2}{\pi}\right)^{1/2}+\left(\theta_e+
\theta_i\right)^{1/2}}{\left(\theta_e+\theta_i \right)^{3/2}},
\ee
where ${\rm ln} \Lambda=20$ is the Coulomb logarithm, and $\theta_i \equiv
kT_i/m_i c^2$ and $\theta_e \equiv kT_e/m_e c^2$ is the dimensionless 
ions and electrons temperatures, respectively. This formula is accurate enough
for our purpose because $q_{ie}$ is only a small
fraction in the energy balance of both ions (eq. 8) and
electrons (eq. 9) (Nakamura et al. 1997). 
$\varepsilon_i$ and $\varepsilon_e$ denote the internal energies for ions
and electrons per unit mass of the gas, respectively. 
Since in ADAFs the ions never
become relativistic while the electrons are transrelativistic, 
following Esin et al. (1997b),
we adopt the following forms\footnote{
based upon the argument of Esin et al.(1997a), 
the adiabatic index $\gamma_{i(e)}$ in Esin et al.(1997b) 
include the contribution of the magnetic density as well. But 
Quataert \& Narayan (1999a) found that this is incorrect if MHD
adequately describes the accretion flow. So we exclude it in the present 
paper. See Quataert \& Narayan (1999b).}:
\be
\varepsilon_i=\frac{1}{\gamma_i-1}\frac{kT_i}{\mu_i m_{\mu}}= 
\frac{3}{2}\frac{kT_i}{\mu_i m_{\mu}},
\ee
\be
\varepsilon_e=\frac{1}{\gamma_e-1}\frac{kT_e}{\mu_e m_{\mu}}
 = a(T_e)\frac{kT_e}{\mu_e m_{\mu}}.
\ee
The expression for 
coefficient $a(T_e)$ is (Chandrasekhar 1939):
\be
a(\theta_e)=\frac{1}{\theta_e}\left[\frac{3K_3(1/\theta_e)+K_1(1/\theta_e)}
{4K_2(1/\theta_e)}-1\right],
\ee
where $K_1, K_2,$ and $K_3$ are modified Bessel 
functions of the second kind of order 1,
2, and 3, respectively.

As for the calculation of the radiative cooling $q^-$, we 
follow the procedure in Manmoto, Mineshige \& Kusunose (1997). 
The considered radiative mechanisms include
bremsstrahlung, synchrotron radiation and Comptonization of 
soft photons. Assuming the disk is isothermal 
in the vertical direction, the spectrum of unscattered photons at a given
radius is calculated by solving the radiative
transfer equation in the vertical direction of the disk basing upon the
two-stream approximation (Rybicki \& Lightman 1979).
The result is (Manmoto, Mineshige \& Kusunose 1997):
\be
 F_{\nu}=\frac{2 \pi}{\sqrt{3}}B_{\nu}[1-{\rm exp}(-2\sqrt{3} \tau^*_{\nu}],
\ee
where $\tau^*_{\nu} \equiv (\pi^{1/2}/2) \kappa_{\nu}H$ is the optical
depth for absorption of the accretion flow in the vertical direction with
$\kappa_{\nu}=\chi_{\nu}/(4\pi B_{\nu})$ being the absorption coefficient,
where $\chi_{\nu}=\chi_{\nu, {\rm brems}}+\chi_{\nu,{\rm synch}}$
is the emissivity, and $\chi_{\nu, {\rm brems}}$ and 
$\chi_{\nu,{\rm synch}}$ are the bremsstrahlung and synchrotron 
emissivities, respectively.  Then the local radiative 
cooling rate $q^-$ reads as follows:
\be
q^-=\frac{1}{2H}\int d \nu \eta(\nu)2F_{\nu},
\ee
where $\eta$ is the energy enhancement factor first introduced by Dermer,
Liang \& Canfield (1991) and modified by Esin et al. (1996)
(See Manmoto, Mineshige \& Kusunose 1997 for the exact formula).

Given the values of parameters $M, \dot{M}, \alpha$ and $\beta$, 
we numerically solve the above set of equations describing the radiation
hydrodynamics of a two-temperature 
 accretion flow around a black hole. The equations are
reduced to a set of differential equations with three variables $v$,
$T_i$ and $T_e$. The global
solutions must satisfy simultaneously the no-torque condition at the horizon,
the sonic point condition at a sonic radius $r_{\rm s}$ 
and three outer boundary conditions given at a certain outer boundary 
$r_{\rm out}$. The numerical method we adopted is the same as in
Nakamura et al.(1997).

We adopt the same procedure as in Manmoto et al. (1997) to calculate 
the emergent spectrum. The spectrum of unscattered photons 
is calculated from equation (14)
and the Compton scattered spectrum is calculated by using the formula 
given by Coppi \& Blandford (1990). Since the Comptonization
is mainly occurred in the inner region of the disk $r \la 10 r_{\rm g}$,
where $H/r \la 0.4$, our ``local'' instead of ``global'' 
treatment of Comptonization
will not cause a serious error.  
 At last, the redshift due to gravity and 
gas motion are also considered in the following simple way
(Manmoto, Mineshige \& Kusunose 1997).
The effect of gravitational redshift is included by taking the ratio
of the the observed energy of a photon to its energy emitted
at radius $r$ to be $(1-r_g/r)^{1/2}$. As for the redshift due to 
the relativistic radial motion, we concentrate 
on the face-on case and take the
rate of energy change to be $1/[1-(v/c)^2]$.

\section{RESULTS}

Throughout this paper, we set $M=10^9 \msun$ and 
$\dot{m}=\dot{M}/\dot{M}_{\rm Edd}=10^{-4}$, where $\dot{M}_{\rm Edd}
=10L_{\rm Edd}/c^2=2.2 \times 10^{-8} M {\rm yr}^{-1}$
is defined as the Eddington accretion rate. Other parameters are
$\alpha=0.1, \beta=0.9$ and $r_{\rm out}=10^3 r_{\rm g}$. 
At $r_{\rm out}$, the three outer boundary conditions 
we imposed are $T_i=T_{{\rm out},i}, T_e=T_{{\rm out},e}$ and 
$\lambda(\equiv v/c_s \equiv v/\sqrt{p/\rho}
)=\lambda_{\rm out}$(see Paper I). 
This set of boundary conditions are equivalent to $(T_{{\rm out},i}, 
T_{{\rm out},e}, \Omega_{\rm out})$ according to eq. (7) 
because $j$ is the eigenvalue of the problem rather 
than  a free parameter. We assign
($T_{{\rm out},i}, T_{{\rm out},e}, \lambda_{\rm out}$)
 to different sets of values 
and investigate their effects on the 
structure of the accretion flow and the
 emergent spectrum. The results are as follows.

We find that the results are qualitatively the same as in Paper I 
although the synchrotron radiation, a strong {\it local} cooling term, 
is included in the electron energy equation. This is because
the differential terms, which 
stand for the {\it global} character of the equations, still play a
dominated role. We find that only when OBCs are 
within a certain range do the global solutions exist.
In terms of the ion temperature, the range is 
$T_{{\rm out},i} \sim (0.01-1) T_{\rm virial}$ ( 
here $T_{\rm virial} \equiv 2GM m_{\mu}
/3kr$ denotes the virial temperature). The electrons temperature is 
slightly lower than $T_{{\rm out},i}$. As for the range of $\Omega_{\rm out}$,
it must satisfy $\Omega_{\rm out} \la 0.8 \Omega_{\rm K}(r_{\rm out})$,
otherwise the viscous heating term takes a negative value under our
viscous description. This result was first pointed out by 
Manmoto, Mineshige \& Kusunose (1997).
The range of OBCs slightly varies under different parameters 
$\dot{m}, \alpha$ and $\beta$, and it is also the function of the value
of $r_{\rm out}$, as we describe in the subsequent section of this paper. 
The structures of the solutions with different OBCs are greatly different.
Three types of solutions 
are also found in terms of their topological structures and
angular momentum profiles (cf. Paper I). 
Generally, when $T_{\rm out}$ is relatively low,
the solution is of type I. When $T_{\rm out}$ is high and 
$\lambda_{\rm out}$ is small (or equivalently $\Omega_{\rm out}$ is large), 
the solution is of type II. Types I and II 
both have small values of sonic 
radii $r_{\rm s} \la 10 r_{\rm g}$, but their angular momentum profiles 
are different (see Figure 1 of Paper I and Figure 1 in the present paper). 
When $T_{\rm out}$ is high and $\lambda_{\rm out}$ is large
(or equivalently $\Omega_{\rm out}$ is
small), the solution becomes type III, which
 has much larger $r_{\rm s}$. 

Figure 1 shows the effect of different $\Ti$ on the solution structures. 
The values of $\Te$ and $\lambda_{\rm out}$ are the same, with $\Te=1.2 
\times 10^8K$ and $\lambda_{\rm out}=0.2$. The solid line 
(belonging to type I solution) 
is for $\Ti=2 \times 10^8K$,
the dotted line (type I) for $\Ti=6 \times 10^8K$,
the short-dashed line (type II) for $\Ti=2 \times 10^9K$ and
the long-dashed line (type III) for $\Ti=3.2 \times 10^9K$.
Six plots in the figure show the radial 
variations of the Mach number (defined as
$v/\sqrt{((3 \gamma_i-1)+2(\gamma_i-1)\alpha^2)/(\gamma_i+1)}/c_s$ so that 
the Mach number equals 1 at the sonic point), the electrons and 
ions temperatures, profiles of the specific angular momentum and 
surface density $\Sigma$ ($\equiv 2\rho H$), and the ratio of the vertical 
scale height of the disk $H(r)$ to the radius, respectively. 
From the figure, we find that the discrepancy 
in the ion temperature is rapidly reduced as the 
radius decreases and $T_i$ converges to
the virial temperature. This is because the local viscous
dissipation in the ions energy equation plays an important
role therefore the temperature of ions is mainly ``locally'' 
determined, like the case of a one-temperature plasma (Paper I).
 However, the discrepancy 
in the surface density of the disk
persists throughout the disk.

Such discrepancy in the solution structures results in the discrepancy
in the emission  spectrum, as Figure 2 shows.
The discrepancy in the luminosity at certain 
individual wave band can be
understood by referring to Figure 1. In the radio-submillimeter band,
the power is principally due to synchrotron
emission for which we approximately have 
$L_{\nu} \propto \rho^{4/5} T_e^{21/5}$ for the ``general''
frequency and $\nu_p L_{\nu_p} \propto 
\rho^{3/2}T_e^7$ for the peak frequency (Mahadevan 1997). Synchrotron radiation
mainly comes from the inner region of the disk, $\la 10r_g$.
So the long-dashed line in the figure possesses the lowest
radio power is mainly due to its lowest surface density, since $T_e$
is almost the same for different solutions. 
For such a low mass accretion rate
system as considered in this paper, the low frequency part of the
submillimeter to hard X-ray spectrum is mainly contributed by
the Comptonization of synchrotron photons, and it is bremsstrahlung emission
that is responsible for the high frequency part.
The long-dashed line still possesses the lowest power in the low frequency band 
 because its corresponding
amount of synchrotron soft photons is the least and 
its corresponding Compton y-parameter $y \equiv \tau_{\rm es}kT_e/m_ec^2$
is the smallest.
In the X-ray band, the solid line has the highest
power. This is because for bremsstrahlung emission we
approximately have $L_{\rm brems} \propto T_e^{-1/2}e^{-h\nu/kT_e} \rho^2$,
i.e., the density instead of $T_e$ is the dominated factor, and the solid
line possesses the highest density among the four solutions.

The effect of $\Te$ on the structures of the global solutions 
is shown in Figure 3. All the three lines in the figure have
$\Ti=2 \times 10^9K$ and $\lambda_{\rm out}=0.2$. 
The solid line is for $\Te=1.2 \times 10^8K$, the dotted line for
$\Te=8 \times 10^8K$ and the dashed line for $\Te=1.2 \times 10^9$.
Different with $T_i$, the discrepancy in $T_e$ persists 
throughout the disk rather than converges as the radius decreases. 
This is because the electrons are essentially adiabatic for the
present low accretion rate case, i.e., both $q_{ie}$ and $q^-$ are very small 
compared with the advection term on the left-hand side of equation (8),
so the electron temperature is globally determined. While for the ion,
the {\em local} viscous dissipation in the energy 
equation plays an important role;
thus, the temperature of ion is determined more locally than the electron.
Such discrepancy in the electron temperature produces significant 
discrepancy in the emission spectrum especially 
in the radio, submillimeter and IR bands,
as Figure 4 shows.

The effect of modifying $\lambda_{\rm out}$ (i.e., modifying $\Omega_{\rm out}$)
on the structure of the accretion flows is shown in Figure 5. 
Each of the three lines in the figure corresponds 
to $T_{{\rm out},i}=2 \times 10^9 {\rm K}$
and $T_{{\rm out},e}=1.2 \times 10^9 {\rm K}$. 
The values of $\lambda_{\rm out}$ are
$0.18$ (solid line, type II), $0.22$ 
(dotted line, type II) and $0.26$ (dashed line, type III), 
and their corresponding angular velocities are 
$0.45 \Omega_{\rm k}, 0.37 \Omega_{\rm k}$ and $0.29 \Omega_{\rm k}$,
respectively. 

When $\Omega_{\rm out}$ decreases across a certain critical
value, $\sim 0.36 \Omega_{\rm k}(r_{\rm out})$ in the present case, 
an accretion pattern characterized by possessing 
a much larger sonic radius appears (referred to as type III solution). 
As we argue later in this paper, this accretion pattern is
of particular interest to us.
In all previous works on viscous accretion flows, only the 
solutions with small sonic radii $\sim $ several $r_{\rm g}$ have been 
found (Muchotreb \& Paczy\'nski 1982; 
Matsumoto et al. 1984; Abramowicz et al. 1988; Chen \& Taam 1993;
Narayan, Kato \& Honma 1997; Chen, Abramowicz \& Lasota 1997; 
Nakamura et al. 1997); therefore, this pattern is new. 
The reason why previous authors did not find this pattern 
is that they generally set the
standard thin disk solutions as the OBCs 
(Muchotreb \& Paczy\'nski 1982; Matsumoto et al. 1984;
Abramowicz et al. 1988; Chen \& Taam 1993; 
Narayan, Kato \& Honma 1997; Chen, Abramowicz \& Lasota 1997),
where the specific angular momentum of the flow is Keplerian;
or, a specific angular momentum with a value near the 
Keplerian one is set as the OBC (Nakamura et al. 1997; 
Manmoto, Mineshige \& Kusunose 1997).

Such transition between the sonic radii happened when 
$\Omega_{\rm out}$ passes across a critical value is clearly shown in 
Figure 6. The value of the critical $\Omega_{\rm out}$
varies with the parameters such as $\alpha$ and $\beta$, and it is 
also the functions of $\Ti, \Te$ and $r_{\rm out}$. 
Generally, it decreases with the decreasing $\alpha$ and/or the 
increasing $r_{\rm out}$. 
Similar transition has been found previously in the context 
of adiabatic (inviscid) accretion 
flow by Abramowicz \& Zurek (1981). In that case,
they found that when the specific angular momentum of the flow $l$, 
a constant of motion, decreased across a critical value $l_c(E)$
(here the specific energy $E$ of the flow is another constant of motion),
a transition from a disk-like accretion pattern to 
a Bondi-like one would happen (Abramowicz \& Zurek 1981;
Lu \& Abramowicz 1988). Here in this paper we find that this 
transition still exist when the flow becomes viscous, confirming 
the prediction of Abramowicz \& Zurek (1981). 

The physical reason of the transition between the two patterns is obvious.
When the specific angular momentum of the gas 
is low, the centrifugal force can be neglected and the gravitational force
plays a dominated role in the radial momentum equation (eq. 6),
therefore the gas becomes supersonic far before the horizon like
in Bondi accretion. When the specific angular momentum is high, 
however, since the centrifugal force becomes stronger, 
the gas becomes transonic only after passing through a sonic 
point near the horizon with the help of the general relativistic 
effect (Abramowicz \& Zurek 1981; Shapiro \& Teukolsky 1984).

Figure 7 shows the corresponding spectra of the solutions 
presented in Figure 5. The accretion flow belonging to the 
new accretion pattern, which is denoted by the
dashed line, emits the lowest X-ray luminosity. This is because
this accretion pattern possesses the largest 
sonic radius therefore the corresponding density 
of the accretion flow is the lowest.

All above results are obtained for a fixed outer boundary $r_{\rm out}=
10^3 r_{\rm g}$. We also investigate the influence of increasing the
value of $r_{\rm out}$ on the global solutions. We find 
that the general feature are quantitatively the same. 
One remarkable difference is that the ranges of $\Ti$ and $\Te$ 
within which we can obtain a global solution lessen with the increasing radii.
For example, if we increase $r_{\rm out}$ from $10^3r_{\rm g}$ to
$10^4r_{\rm g}$, the temperature
range will lessen from $\sim (0.01-1) T_{\rm virial}$ 
to $\sim (0.1-1) T_{\rm virial}$. If $r_{\rm out}$ is taken to be large enough,
$\Ti (\approx \Te)$ almost be unique.
However, the feasible range of $\Omega_{\rm out}$
is almost constant, no matter how large $r_{\rm out}$ is. 
Moreover, when the value of $r_{\rm out}$ becomes larger,
the transition between the accretion patterns, happened 
when $\Omega_{\rm out}$ pass across the critical value, becomes more ``obvious''
in the sense that the small sonic radius becomes 
smaller and the large sonic radius becomes 
larger. Consequently, the discrepancy between the surface density 
of the accretion disk of the two accretion patterns becomes 
larger compared with the case of small $r_{\rm out}$, and this will further 
result in the increase of the discrepancy of the X-ray 
luminosity emitted by the accretion flow. This is the crucial factor
to solve the puzzle of the mass accretion rate of Sgr A$^*$.

\section{APPLICATION TO Sgr A$^*$}

Kinematic measurements suggest that the energetic radio source
Sgr A$^*$ located at the center our Galaxy
 is a supermassive compact object with a mass
$\sim 2-3 \times 10^6 \msun$.
 This is widely believed to be a black hole (Mezger, Duschl \& Zylka 1996).
On the other hand, observations of gas outflows
near Sgr A$^*$ indicate the existence of
a hypersonic stellar wind coming from
the cluster of stars within several arc-seconds from Sgr A$^*$.
The wind should be accreting onto the black hole (Melia 1992).
If the flow past Sgr A$^*$ is uniform, then
the mass accretion rate can be simply obtained by
the classical Bondi-Hoyle scenario (Bondi \& Hoyle 1944) as follows.
Since the wind is significantly hypersonic, as such, a standing bow-shock
is inevitable. This shock is located roughly where flow elements'
potential energy equals kinetic energy (Shapiro 1973; Melia 1992),
$R_{\rm A}=2GM_{\rm BH}/v^2_w$,
where $M_{\rm BH}$ is the mass of the black hole and $v_w$ is
the wind velocity. From here the shocked gas is assumed to
accret into the hole. Since the mass flux in the gas at large radii is
$m_p n_w v_w$, where $m_p$ is proton mass and $n_w$ is the number density
of the gas, the mass accretion rate is
$\dot{M}_{\rm BH}=\pi R_{\rm A}^2 m_p n_w v_w$.

This result is only valid for an uniform source.
In reality the flow past Sgr A$^*$ comes from multiple sources.
In this case, the wind-wind shocks dissipate some of
the bulk kinetic energy and lead to a higher capture rate for
the gas (Coker \& Melia 1997).
The exact value of accretion rate $\dot{M}$ depends on the stellar spatial
distribution and can only be obtained by numerical simulation.
The three-dimensional hydrodynamical simulation of
Coker \& Melia (1997) gives the average value $\dot{M}=1.6 \dot{M}_{\rm BH}$
for two extreme spatial distributions, spherical and planar ones.
With the available data at that time,
the accretion rate obtained in Coker \& Melia (1997)
is $\dot{M} \sim 10^{22} \gs$.
Considering that recent work suggests
the Galactic Centre wind is dominated by a few hot stars with
higher wind velocities of $\sim 1000 {\rm \,km}\,s^{-1}$(Najarro et al. 1997)
rather than $700 {\rm \,km \,s}^{-1}$ taken by Coker \& Melia (1997),
the more reliable accretion rate should be
$\dot{M} \sim (700/1000)^3 \times 10^{22} \gs \sim
9 \times 10^{-4} \dot{M}_{\rm Edd}$.

Assuming the accretion is via a standard thin disk,
the accretion rate required to model the luminosity
is more than three orders of magnitude
lower than this value. 
ADAF model has been turned out to be
a significant success to model its low luminosity 
(Narayan, Yi \& Mahadevan 1995; Manmoto, Mineshige \&
Kusunose 1997; Narayan et al. 1998).
However, there exists a discrepancy between the accretion rate favored 
by all ADAF models
in the literature and that favored by the simulation mentioned above, 
with the former being 10-20 times smaller than that favored by the 
latter (Coker \& Melia 1997; Quataert \& Narayan 1999b).
The most up-to-date calculation
gives $\dot{M}=6.8 \times 10^{-5} 
\dot{M}_{\rm Edd}$ (Quataert \& Narayan 1999b) and this value 
approaches the lower limit considered plausible 
from Bondi capture (Quataert, Narayan, \& Reid 1999). 
In an ADAF model, the mass accretion
rate is determined by fitting the theoretical
X-ray flux to the observation.
If a larger accretion rate
were adopted, then bremsstrahlung radiation
would yield an X-ray flux well
above the observational limits. Even though we assume that significant
accretion mass may be lost to a wind, detailed calculation shows
that since the bremsstrahlung radiation comes from large radii in the
accretion flow, the discrepancy can not be
alleviated no matter how strong the winds are (Quataert \& Narayan 1999b).

The present research on the role of OBC tells us that we should 
seriously consider the physical state of accretion flows
at the outer boundary. However, we note that
in all present ADAF models of Sgr A$^*$,
 the outer boundary condition
is roughly treated. In our view, this might be an origin of the discrepancy
between the mass accretion rate.

Basing upon the above consideration,
we recalculate the spectrum of Sgr A$^*$.
We set the outer boundary at the
``accretion radius'' $R_A=2GM/v_w^2 \sim 1.5 \times
10^5r_{\rm g}$. In the present case of Sgr A$^*$, 
the temperatures of ions and electrons
in the flow just after the bow-shock should equal the virial temperature
$\sim 10^{12}/(r/r_{\rm g})K$.
But the value of the specific angular momentum at the outer boundary
is not certain. We set it as various values and find that,
when it pass across a critical value $\sim 0.16 \Omega_{\rm K}$,
the transition of the accretion pattern
occurs. Figure 8 shows the Mach number and the surface density of
the two accretion patterns.
The parameters of both the solid and the dashed lines
in the figure are: 
$M_{\rm BH}=2.5 \times 10^6 \msun$,
$\dot{M}=4 \times 10^{-4} \dot{M}_{\rm Edd}$,
$\alpha=0.1$ and $\beta=0.9$. 
At the outer boundary $R_{\rm A}$, the two lines possess identical
temperature of $8 \times 10^6{\rm K}$, but their specific angular momenta
are different. The solid line, which stands for our new accretion
pattern, corresponds to relatively
low angular velocity, $\sim 0.15 \Omega_{\rm K}$, while 
the angular velocity possessed by the dashed 
line, which we draw for comparison,
is much higher, $\sim 0.46 \Omega_{\rm K}$.
Their sonic radii are $\sim 4 r_{\rm g}$ (dashed line),
and $\sim 6000 r_{\rm g}$ (solid line), respectively.
The difference
of the sonic radius results in the discrepancy in the surface density,
which further results in the difference of the X-ray luminosity, 
as shown by Figure 9. 

Figure 9 shows our calculated X-ray spectra
together with the observation of Sgr A$^*$. Here we assume that the
bremsstrahlung radiation is the only contributor to this waveband.
This assumption requires that the synchrotron radiation
can not be too strong, otherwise the contribution from the
Comptonization of the synchrotron photons
will exceed that from bremsstrahlung radiation.
This requirement can be satisfied according to 
the radio observation of Sgr A$^*$.
We will discuss this question in the following part.
From the figure, we find that accretion rate as high as
$4 \times 10^{-4} \dot{M}_{\rm Edd}$ is
acceptable, if only the angular momentum
at the outer boundary is relatively low. Compared with
the value of $6.8 \times 10^{-5} \dot{M}_{\rm Edd}$
in Quataert \& Narayan (1999b), this value is much
closer to that favored by the numerical
simulation. If the angular momentum of the flow at $R_{\rm A}$
is relatively high, however, as shown by the dashed line, the X-ray
flux is well above the observation.

Now the crucial moment is whether the value
of the angular velocity at $R_{\rm A}$ is really low.
In this context, we note that the three-dimensional
numerical simulation by Coker \& Melia (1997) indicates that
the accreted angular velocity
in Sgr A$^*$ is very low, $\Omega_{\rm out} \approx
0.1 \Omega_{\rm K}(r_{\rm out})$ for their run 1 and
$\approx 0.2 \Omega_{\rm K}(r_{\rm out})$ for their run 2.
The consistency with our result is satisfactory.

\section{SUMMARY AND DISCUSSION}

In this paper, we numerically solve the radiation hydrodynamic
equations describing an optically thin advection dominated accretion flow
and calculate its emergent spectrum.
We fix all the parameters such as $\alpha, \beta, M$ and
$\dot{M}$, and set the outer boundary condition as $\Ti, \Te$
and $\lambda_{\rm out}(\equiv v/c_s)$
at certain outer boundary $r_{\rm out}$.
Our primary focus is to investigate the effects of the outer boundary
conditions on the structure of the global
solution and, especially, on the emergent spectrum.

We find that OBCs must lie in a certain range,
otherwise we can't find the corresponding global solutions.
However, this range is large enough in the sense that both the structure
of the solution and its corresponding spectrum differ greatly
from each other under different OBCs. 
Three types of solutions
are also found under various OBCs,
qualitatively agree with the result of Paper I, where magnetic
field is neglected and the description to radiation is crude.
The value of the sonic radius in type III solution
is much larger than types I and II.  We examine the
individual influence of $\Ti, \Te$ and $\lambda_{\rm out}$ on
the structures and spectrum and find that each factor especially
the temperature, plays a significant role. The peak flux in
the radio, IR and X-ray  bands can differ by nearly two 
orders of magnitude or more in our example, 
although all the other parameters are exactly the same.

We also investigate the effect of modifying the value of $r_{\rm out}$.
We find that the feasible ranges of $\Ti$ and $\Te$ lessen with the
increasing radii. If the value of $r_{\rm out}$ is too large, the values of
$\Ti$ and $\Te$ are almost unique. 
This result seems to mean the reduction of the effect of OBC on the solution.
However, on the one hand we expect that 
in some cases, such as some binary system
or the system where a thin disk-ADAF transition is 
expected to occur (e.g., X-ray binaries A0620-00 
and V404 Cyg, and low luminosity AGN NGC 4258; see Narayan, Mahadevan, \&
Quataert 1998 and references therein), 
 the value of $r_{\rm out}$
might be small. On the other hand, the feasible range of 
$\Omega_{\rm out}$ is almost constant no matter how large 
$r_{\rm out}$ is.  

The reason why previous works on ADAF global solution
(e.g., Narayan, Kato, \& Honma 1997; Chen,
Abramowicz, \& Lasota 1997)
did not find the obvious effect of OBC on the global solution 
is that, both Narayan, Kato \&
Honma (1997) and Chen, Abramowicz \& Lasota (1997) concentrated
on the dynamics of a {\em one-temperature plasma}, 
where the {\em local} viscous dissipation 
in the energy equation plays an important role. 
Due to this reason, their global solutions are in principle 
``locally'' rather than ``globally'' 
determined, the effect of OBC weakens rapidly 
with the decreasing radii and the solutions converged rapidly over
a small radial extent away from the outer boundary. 
We also obtained similar results with theirs
for the one-temperature plasma in our Paper I. 
However, according to our result, the variation of $\Omega_{\rm out}$ 
across a certain critical value
will produce obvious OBC-dependent behavior such as the 
transition of the sonic radius,
no matter what type the plasma is, one- or two-temperature. 
They did not find this result might be due to the 
fact that  $\Omega_{\rm out}$ adopted by them, 
$\Omega_{\rm K}(r_{\rm out})$ for Keplerian disk 
outer boundary condition or
$\sim 0.34 \Omega_{\rm K}(r_{\rm out})$ 
for self-similar solution outer boundary condition, is 
always larger than the critical value, so the effect of OBC
is very small hence is hard to find. 
 
The present study concentrates on the low-$\dot{M}$ case where
the differential terms in the equation such as the energy advection
play an important role therefore the effect of OBC are most obvious.
When $\dot{M}$ becomes higher, the role of 
the local radiation loss terms in the energy balance
will become more important. In this case, we expect that the discrepancy 
due to OBC in the profile of the 
temperature will lessen. However, from our calculation to
the one-temperature accretion flow whose temperature
is also principally determined locally (Paper I), and Figure 5 in 
the present paper, we expect
that the flow should still present OBC-dependent
behavior in, e.g., the angular momentum and the Mach number
profiles. 

We do not include winds in the present study. Since the Bernoulli parameter
of ADAF is positive therefore the gas can in principle 
escape to infinity with positive energy (Narayan \& Yi 1994, 1995).
Blandford \& Begelman (1999) recently suggested that mass loss
through winds might be dynamically important. The effect of winds on the
spectrum of ADAF has been investigated by Quataert \& Narayan (1999b).
It is interesting to investigate the effect 
of OBC on ADAF with winds. We expect the result is probably similar
with the present results since in that case the differential terms in 
the equation still play an important role.

It is a meaningful problem whether a standing shock occurs
in an accretion flow. Although some authors have set up
the shock-included global solutions (e.g. Chakrabarti 1996),
the result is generally thought not to be so
convincing because in their numerical procedure
$r_{\rm s}$ and $j$ are treated as two
free parameters instead of
the eigenvalues of the problem.
A necessary condition for the shock
formation is the existence of the global solution
with a large sonic radius outside the centrifugal 
barrier. This is the key of the problem.
According to our result, this kind of large-sonic-radius 
solution, belonging to our type III, can only be 
realized when the specific angular momentum of the accretion flow
is low. We note that this requirement to the angular momentum
is exactly what Narayan, Kato \& Honma(1997)
anticipated, and it has recently been
confirmed by the numerical simulation by Igumenshchev, Illarionov \&
Abramowicz (1999).

The large-sonic radius solution (type III) is a new accretion pattern since 
in all previous studies on viscous accretion onto black holes 
the sonic radii are small. We find that the discrepancy in the mass
accretion rate of Sgr A$^*$ between the value favored by the previous 
ADAF models in the literature and that favored by the 
hydrodynamical numerical simulation is significantly reduced if the accretion
is via this pattern. This result hints us that
such low angular momentum accretion may be
very common in the universe. One example is the detached binary system,
where the accretion material is the stellar wind from the companion
therefore the angular momentum of the accreting gas is
very low (Illarionov \& Sunyaev 1975). 
Another example is
the cores of nearby giant elliptical galaxies, where the angular momentum
of the hot (accretion) gas is again assumed to be 
very small.

\acknowledgements

We are grateful to the referee for 
many helpful comments and suggestions which enabled us 
to improve the presentation. 
This work is supported in part by the National Natural Science
Foundation of China under grant 19873007. F.Y. also
thanks the financial support
from China Postdoctoral Science Foundation.

\references
\def\refpar{\hangindent=3em\hangafter=1}
\def\reference{\refpar\noindent}
\def\apj{ApJ}
\def\apjs{ApJS}
\def\mnras{MNRAS}
\def\aa{A\&A}
\def\aas{A\&A Suppl. Ser.}
\def\aj{AJ}
\def\araa{ARA\&A}
\def\nat{Nature}
\def\pasj{PASJ}

\reference Abramowicz, M.A., Chen, X., Kato, S., Lasota, J.-P., \&
Regev, O. 1995, \apj, 438, L37

\reference Abramowicz, M.A., Czerny, B., Lasota, J.P., \& Szuszkiewicz, E.,
1988, \apj, 332, 646

\reference Abramowicz, M., A., Zurek, W.H., 1981, ApJ, 246, 31

\reference Begelman, M.C., 1978, \mnras, 243, 610

\reference Begelman, M.C., \& Meier, D.L., 1982, \apj, 253, 873

\reference Blandford, R.D., \& Begelman, M.C. 1999, MNRAS, 303, L1

\reference Bondi, H. \& Hoyle, E. 1944, \mnras, 104, 273

\reference Chakrabarti, S.K. 1996, \apj, 464, 664

\reference Chandrasekhar, S. 1939, Introduction to the Study of
Stellar Structure(New York: Dover)

\reference Chen, X., Abramowicz, M. A., \& Lasota, J.-P. 1997, \apj, 476, 61

\reference Chen, X. \& Taam, R. 1993, \apj, 412, 254
 
\reference Coker, R., \& Melia, F., 1997, \apj, 488, L149

\reference Coppi, P.S., \& Blandford, R.D. 1990, \mnras, 245, 453

\reference Dermer, C.D., Liang, E.P., \& Canfield, E. 1991, \apj, 369, 410

\reference Esin, A.A., Narayan, R., Ostriker, E., \& Yi, I. 1996, \apj, 465, 312

\reference Esin, A.A., 1997a, \apj, 482, 400

\reference Esin, A.A., McClintock, J. E., \& Narayan, R. 1997b, \apj, 489, 865

\reference Ichimaru, S., 1977, \apj, 214, 840

\reference Igumenshchev, I.V., Illarionov, A.F. \& Abramowicz, M.A. 1999, 
\apj, 517, L55

\reference Illarionov, A.F., \& Sunyaev, R.A., 1975, \aa, 39, 185

\reference Lu, J.F., \& Abramowicz, M.A. 1988,  Acta Ap. Sin., 8, 1
 
\reference Mahadevan, R. 1997, \apj, 477, 585

\reference Manmoto, T., Mineshige, S., Kusunose, M. 1997, \apj, 489, 791

\reference Matsumoto, R., Kato, S., Fukue, J., \& Okazaki, A.T. 1984, \pasj, 36,
71

\reference Melia, F 1992, \apj, 1992, 387, L25

\reference Mezger, P.G., Duschl, W.J., \& Zylka, R. 1996, A\&AR, 7, 289

\reference Muchotreb, B., \& Paczy\'nski, B. 1982, Acta Astron. 32, 1

\reference Najarro, E., et al. 1997, \aa, 325, 700

\reference Nakamura, K. E., Kusunose, M., Matsumoto, R., \& Kato, S. 1997,
\pasj,  49, 503

\reference Narayan, R., Kato, S. \& Honma, F. 1997, \apj, 476, 49

\reference Narayan, R., Mahadevan, R., Grindlay, J.E., Popham, R. \&
 Gammie, C., 1998, \apj, 492, 554

\reference Narayan, R., Mahadevan, R., \& Quataert, E. 1998, in ``The Theory
of Black Hole Accretion Discs'', eds. M.A. Abramowicz, G. Bjornsson,
and J.E. Pringle, (Cambridge University Press)

\reference Narayan, R. \& Yi, I. 1994, \apj, 428, L13

\reference Narayan, R. \& Yi, I. 1995, \apj, 444, 231

\reference Narayan, R., Yi, I., Mahadevan, R. 1995, \nat, 374, 623

\reference Paczy\'nski, B., \& Wiita, P. J. 1980, \aa, 88, 23

\reference Quataert, E., \& Narayan, R., 1999a, \apj, 516, 399

\reference Quataert, E., \& Narayan, R., 1999b, \apj, 520, 298 

\reference Quataert, E., Narayan, R., \& Reid, M.J. 1999, \apj, 517, L101

\reference Rybicki, G., \& Lightman, A.P. 1979, Radiative Processes 
in Astrophysics (New York: Wiley)

\reference Rees, M.J., Begelman, M.C., Blandford, R.D., \& Phinney, E.S., 
1982, \nat, 295, 17

\reference Shakura, N.I., \& Sunyaev, R.A., 1973, \aa, 24, 337

\reference Shapiro, S.L. 1973, \apj, 180, 531

\reference Shapiro, S.L., \& Teukolsky, S.A. 1984, 
{\em Black Holes, White Dwarfs,
and Neutron Stars} (New York: Wiley)

\reference Yuan, F. 1999, \apj, 521, L55 (Paper I)

\newpage

\begin{figure}
\psfig{file=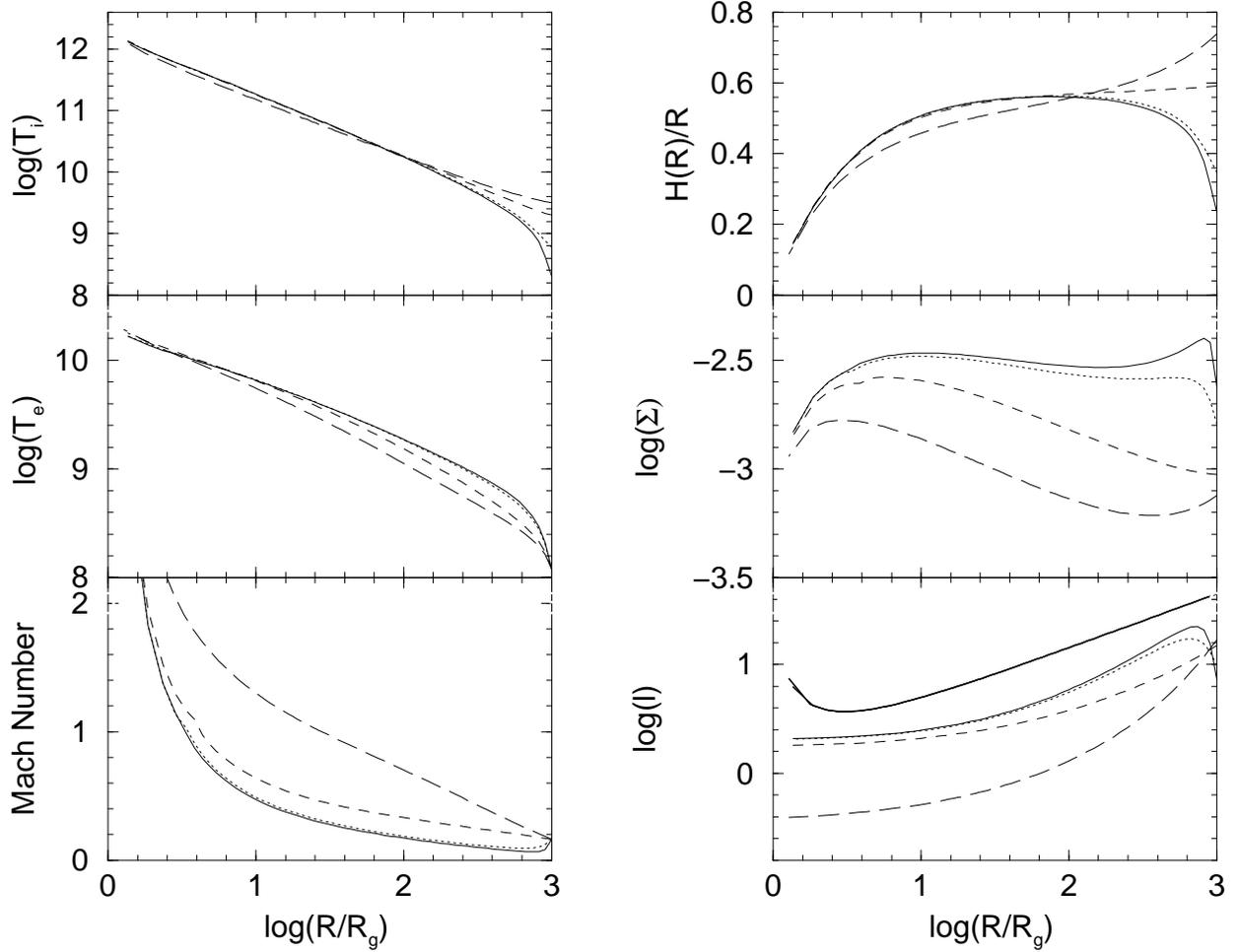,width=1.\textwidth,angle=270}
\caption{The structures of the accretion flow with
different $\Ti$. The solid line (type I solution)
is for $\Ti=2 \times 10^8K$,
the dotted line (type I) for $\Ti=6 \times 10^8K$,
the dashed line (type II) for $\Ti=
2 \times 10^9K$ and the long-dashed line (type III)
for $\Ti=3.2 \times 10^9K$.
Other OBCs are $\Te=1.2 \times 10^8K$ and $\lambda_{\rm out}=0.2$.
The outer boundary is set at $r_{\rm out}=10^3r_{\rm g}$.
Other parameters are
$\alpha=0.1, \beta=0.9, M=10^9 \msun$ and $\dot{M}=10^{-4} \dot{M}_{\rm Edd}$.
The units of $\Sigma$ and $T$ are  ${\rm g \ cm^{-2}}$ and K}
\end{figure}

\begin{figure}
\psfig{file=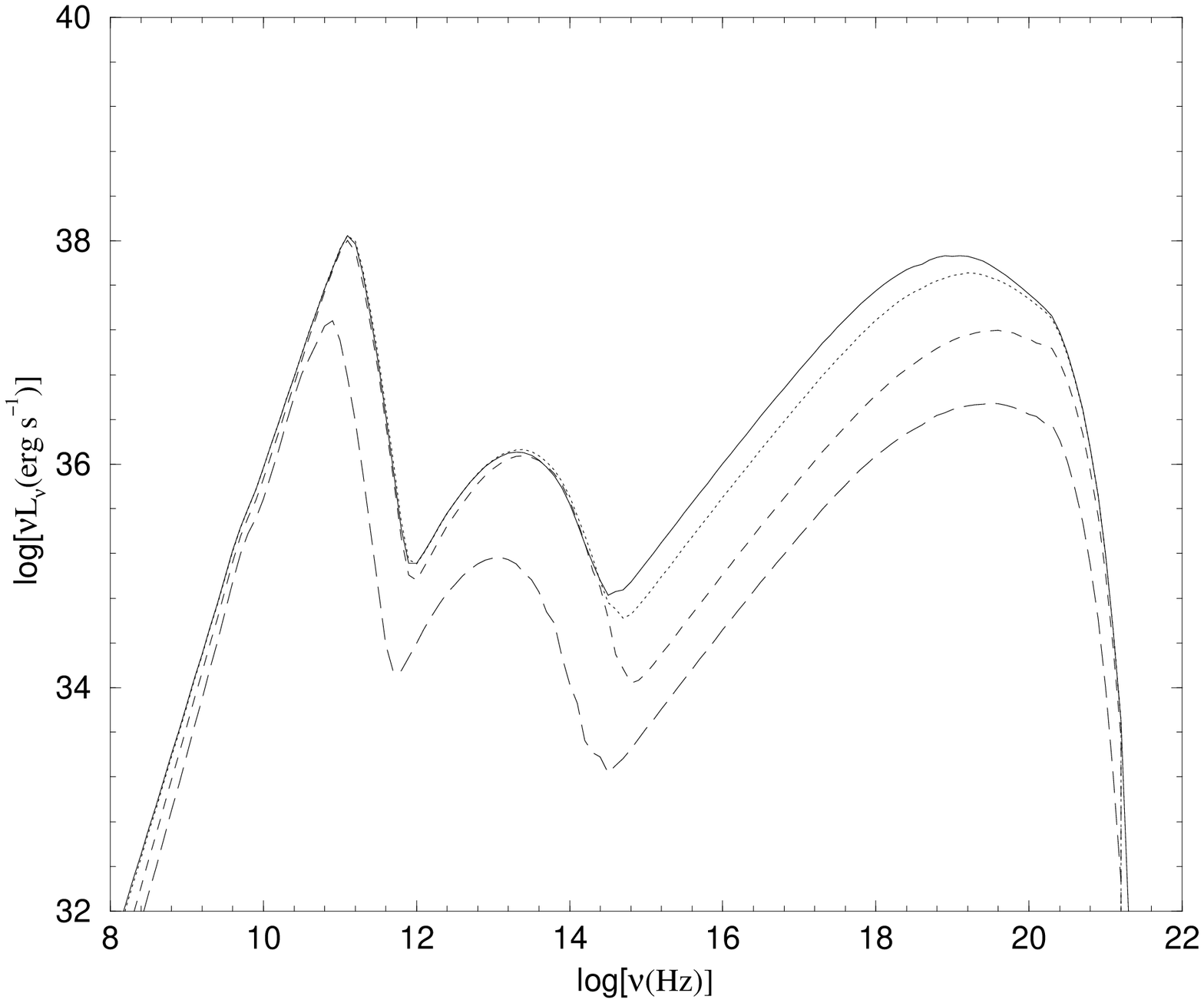,width=1.\textwidth,angle=270}
\caption{The corresponding spectra of the solutions shown in Figure 1.}
\end{figure}

\begin{figure}
\psfig{file=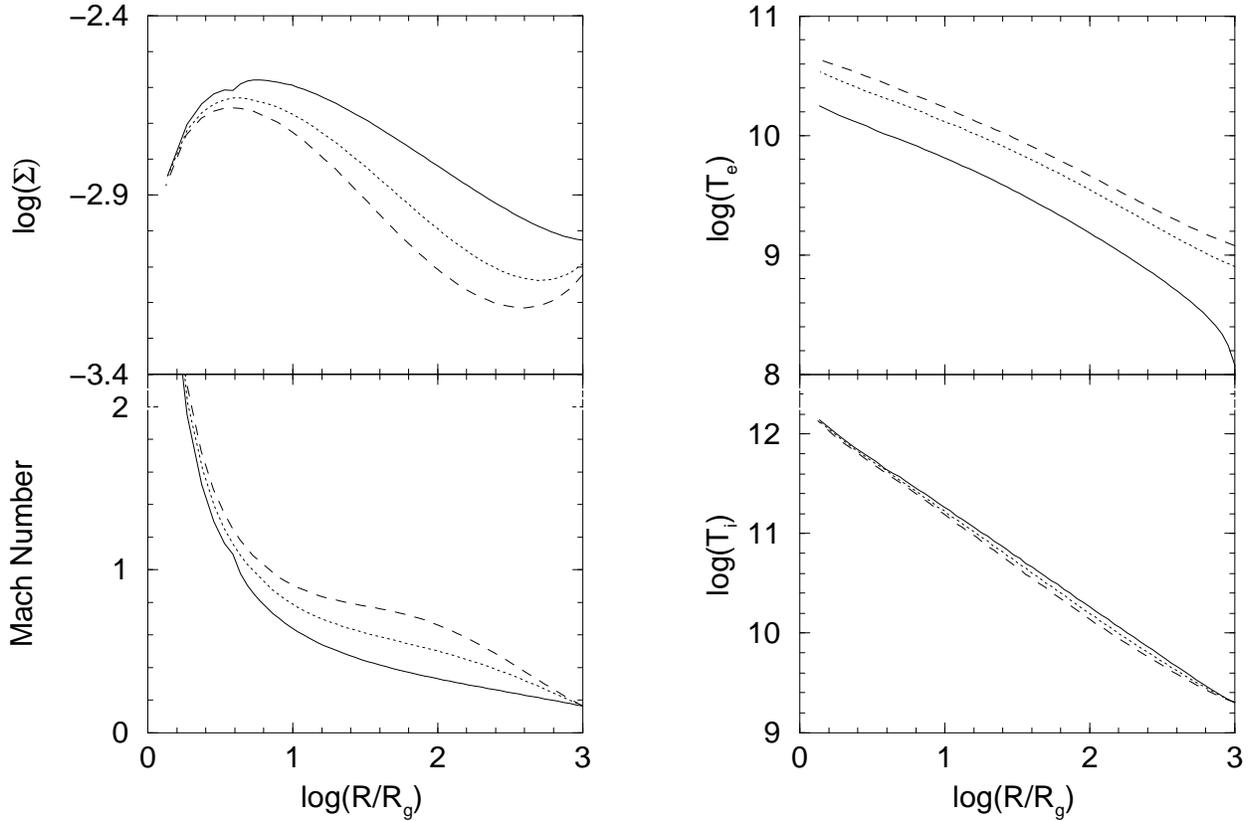,width=1.\textwidth,angle=270}
\caption{The structures of the accretion flows with different $\Te$.
The solid, dotted and the dashed lines are for $\Te=1.2 \times 10^8K,
8 \times 10^8 K$ and $1.2 \times 10^9K$, respectively.
Other OBCs are $\Ti=2 \times 10^9K$ and $\lambda_{\rm out}=0.2$.
The outer boundary is set at $r_{\rm out}=10^3r_{\rm g}$.
Other parameters and the units are the same as those in Figure 1.}
\end{figure}

\begin{figure}
\psfig{file=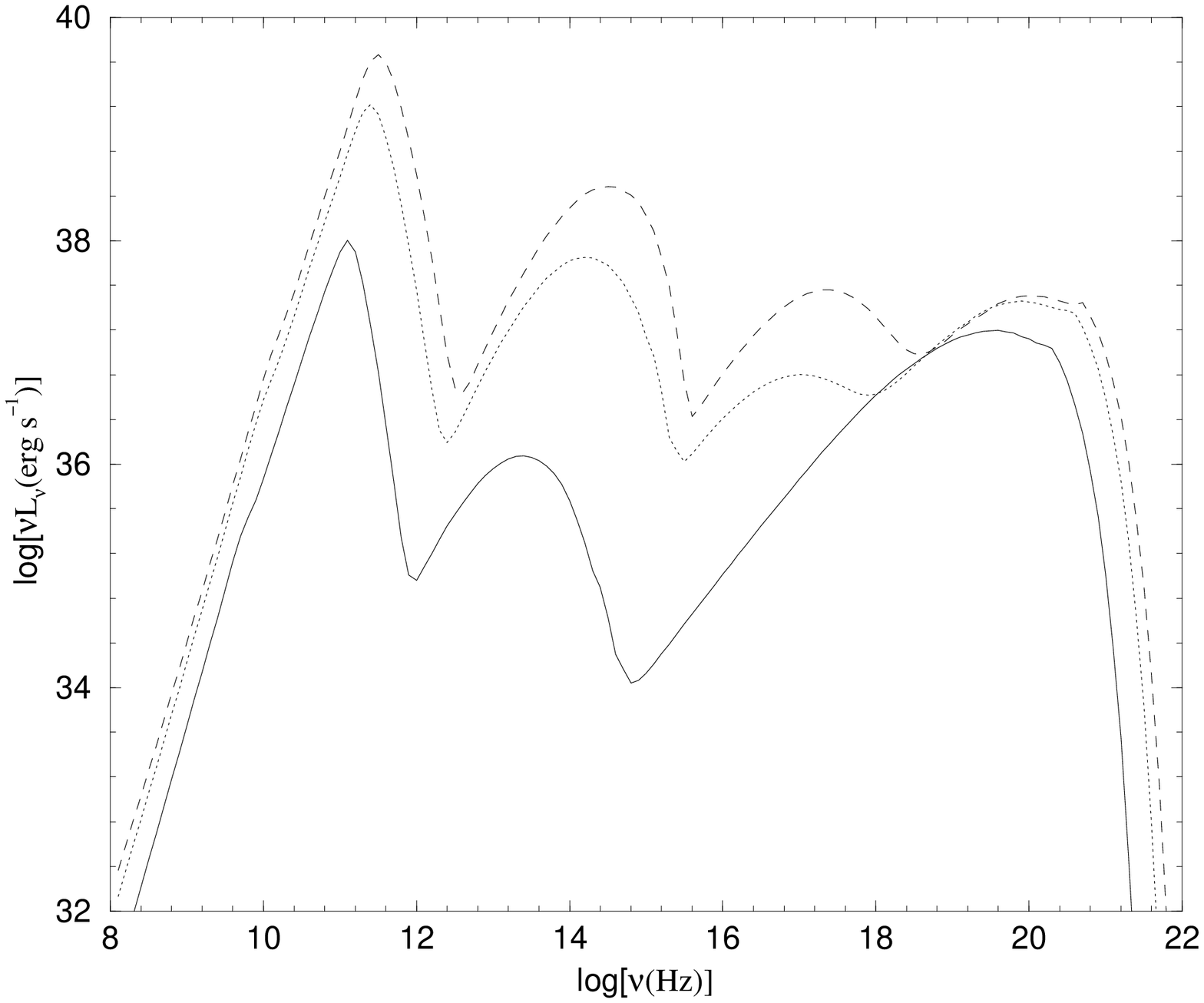,width=1.\textwidth,angle=270}
\caption{The corresponding spectra of the solutions shown in Figure 3.}
\end{figure}

\begin{figure}
\psfig{file=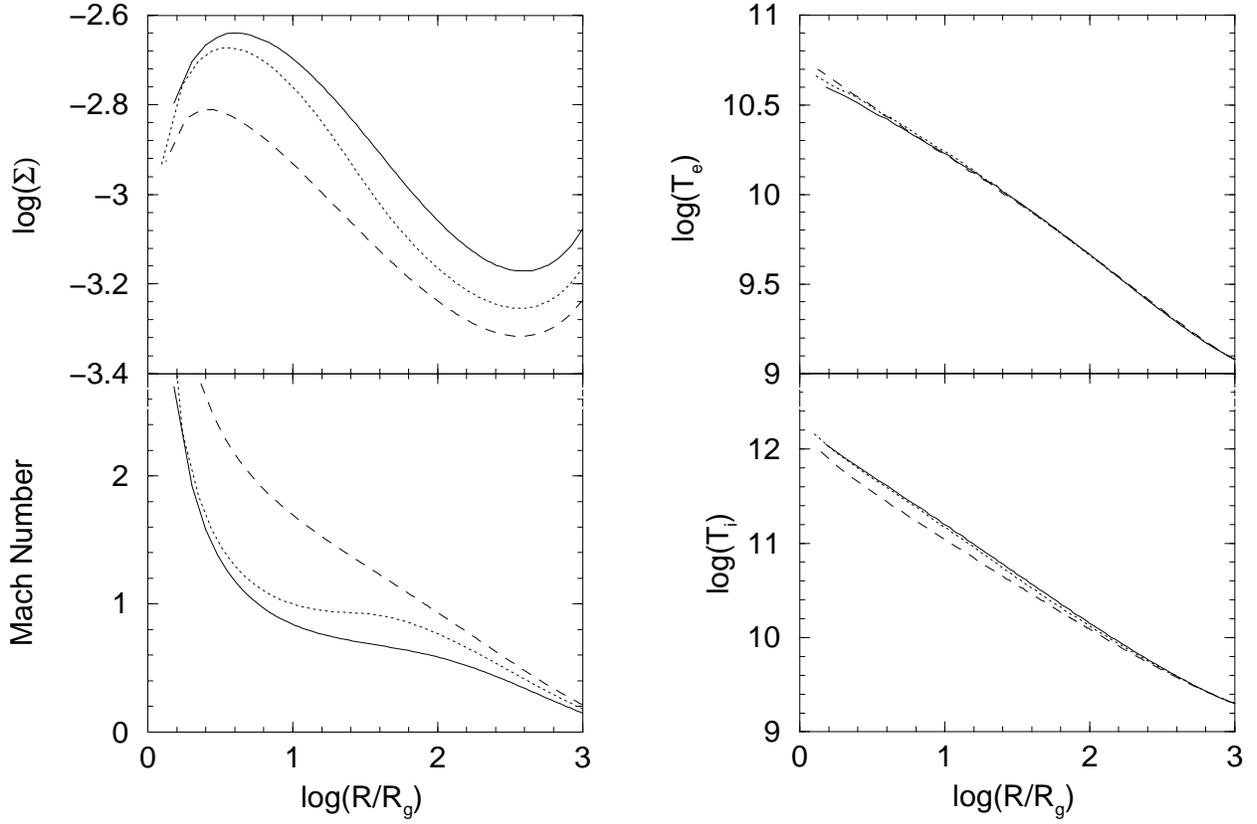,width=1.\textwidth,angle=270}
\caption{The structures of the accretion flows with
different $\lambda_{\rm out}$.
The solid, dotted and the dashed lines are for
$\lambda_{\rm out}=0.18,
0.22$ and $0.26$, respectively. The corresponding angular velocities are
$0.447 \Omega_{\rm K}, 0.37 \Omega_{\rm K}$ and
$0.289 \Omega_{\rm K}$.
Other OBCs are $\Ti=2 \times 10^9K$ and $\Te=1.2 \times 10^9K$.
The outer boundary is set at $r_{\rm out}=10^3r_{\rm g}$.
Other parameters and the units are the same as those in Figure 1.}
\end{figure}

\begin{figure}
\psfig{file=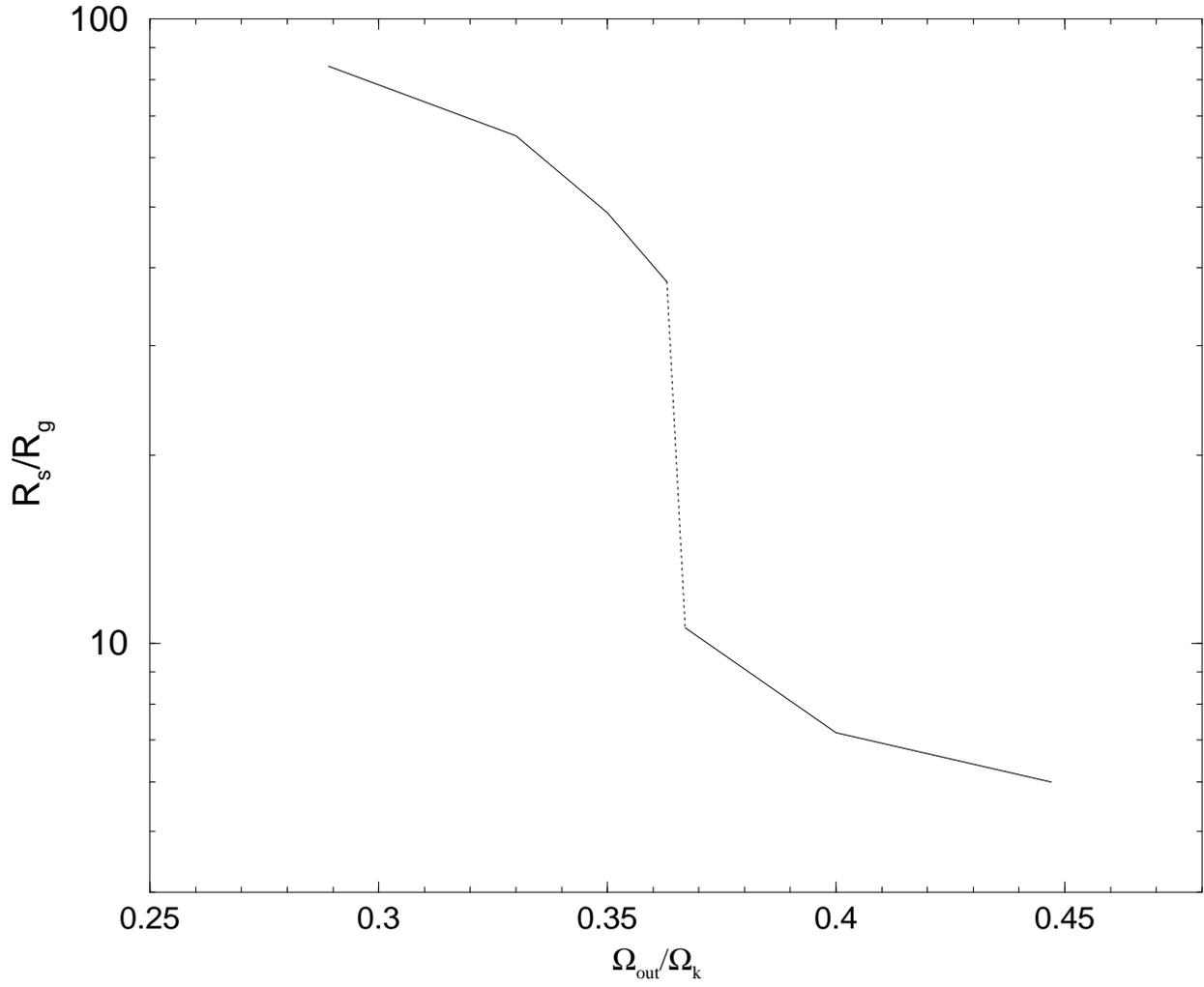,width=1.\textwidth,angle=270}
\caption{The variation  of the value of the sonic radii with
the angular velocity at the outer boundary. A transition is clearly shown.}
\end{figure}

\begin{figure}
\psfig{file=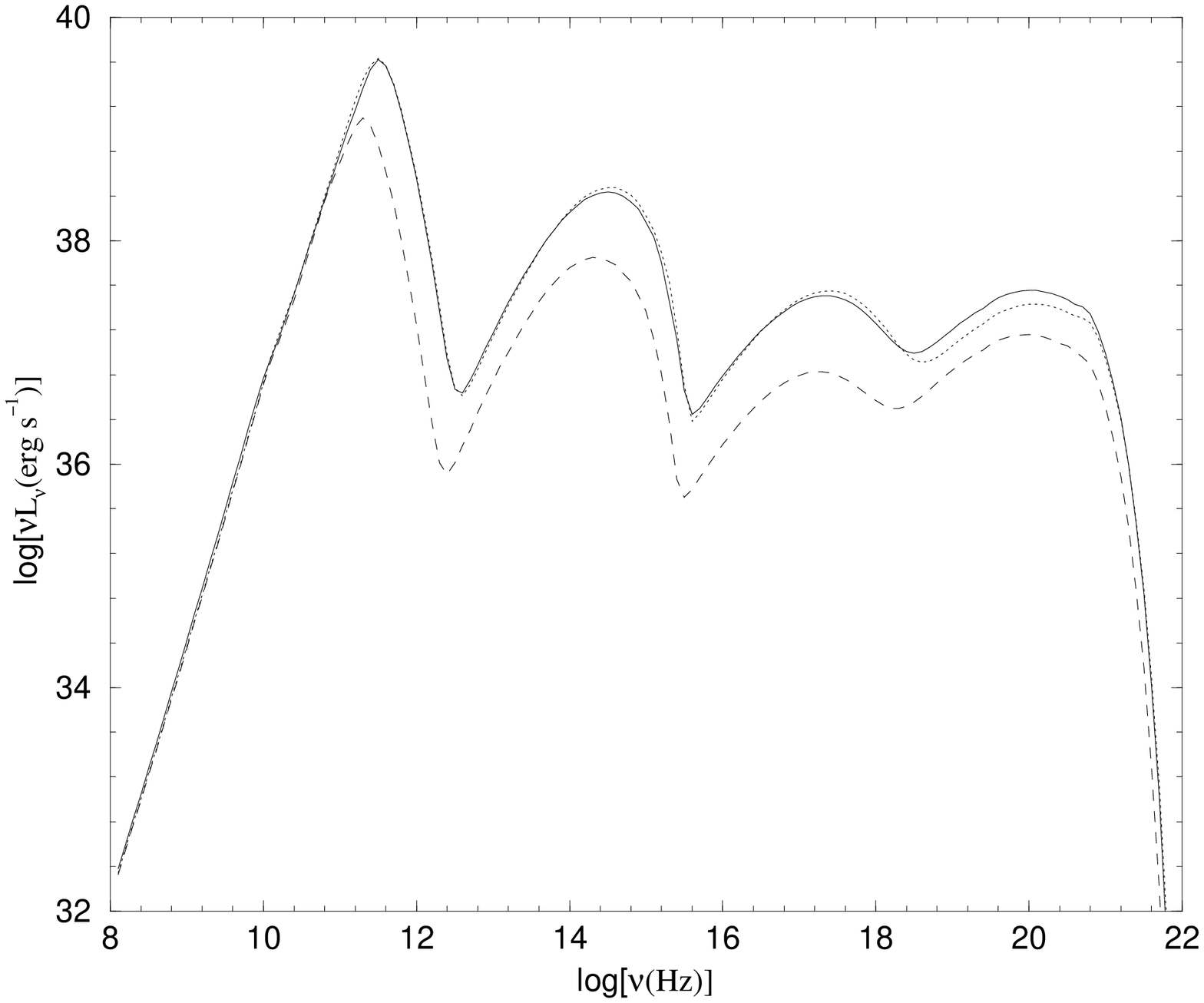,width=1.\textwidth,angle=270}
\caption{The corresponding spectra of the solutions shown in Figure 5.}
\end{figure}

\begin{figure}
\psfig{file=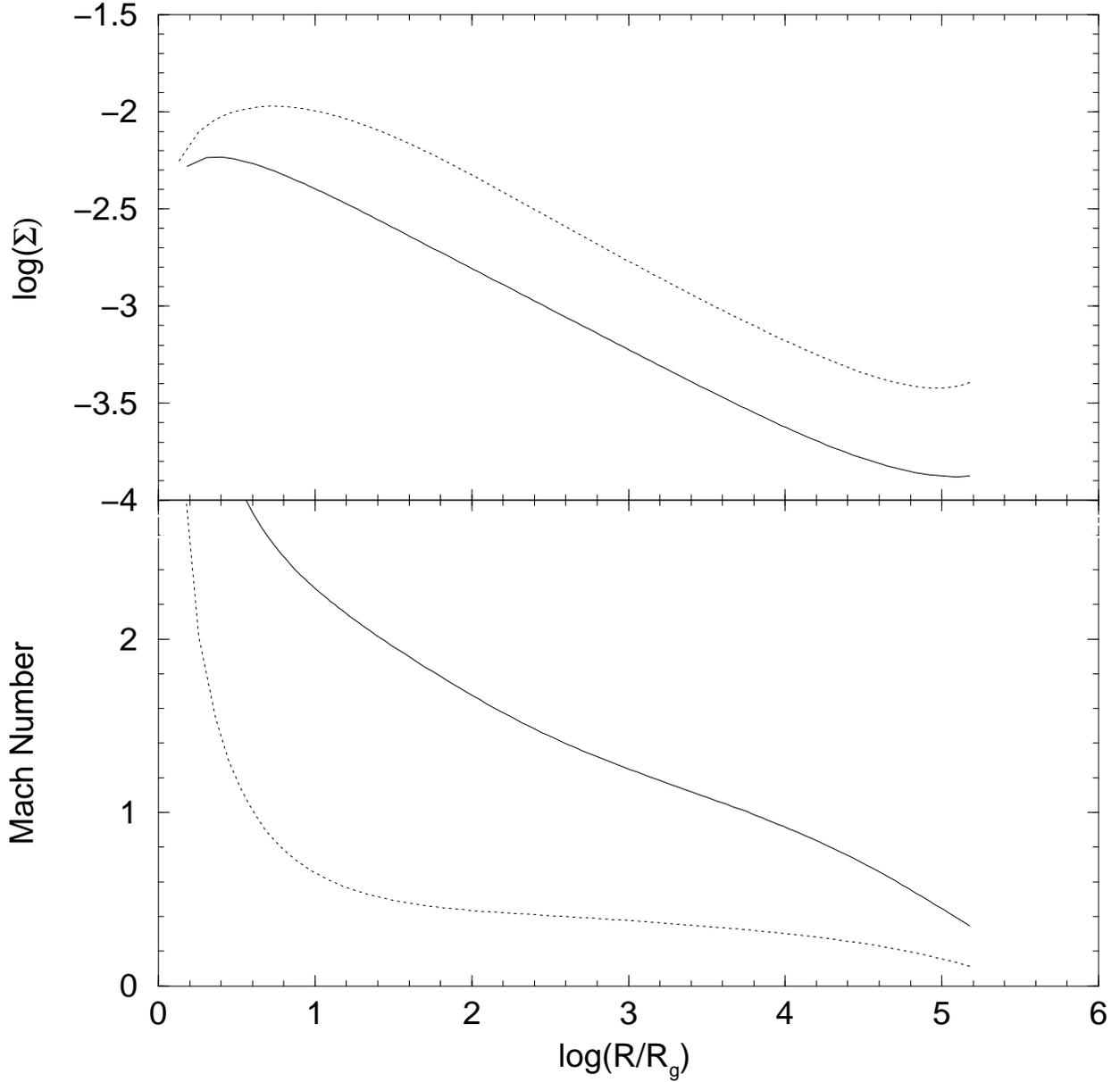,width=1.\textwidth,angle=270}
\caption{The radial variations of the
surface density and the Mach
number of the accretion flows with different
specific angular momenta at the outer
boundary $R_{\rm A}$. The solid line, which
stands for our new accretion pattern,  is for
$\Omega_{\rm out}=0.15 \Omega_{\rm K}$ while the dashed line
is for $\Omega_{\rm out}=0.46 \Omega_{rm K}$. }
\end{figure}

\begin{figure}
\psfig{file=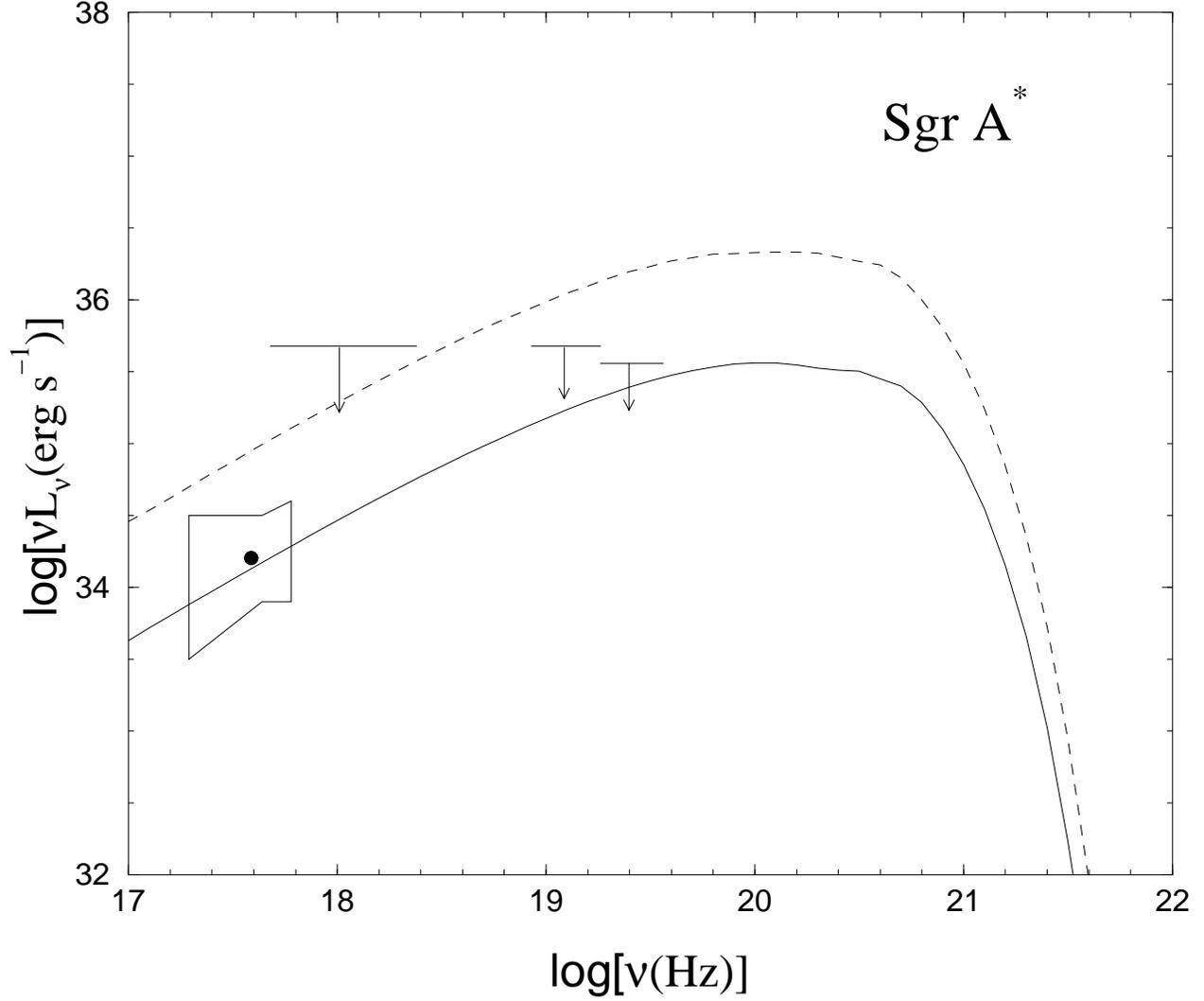,width=1.\textwidth,angle=270}
\caption{The X-ray spectrum of Sgr A$^*$.
The measured fluxes were converted to luminosity  assuming
a distance of 8.5 kpc to the Galactic Centre. The observational
data are compiled by Narayan et al. (1998).
The spectra represented by the solid
and the dashed lines are produced by the accretion flows denoted
by the same style of lines as in Figure 8.
Due to the difference of the angular momentum of the flow at the outer
boundary, the X-ray flux differs by a factor $\sim$ 8.}
\end{figure}

\end{document}